%% file: mc2021_latex.tex
\documentclass[letterpaper]{mc2021}
\usepackage{tabls}
\usepackage{cites}
\usepackage{epsf}
\usepackage{ragged2e}
\usepackage{amsmath}
\usepackage{amssymb}
\usepackage[english]{babel}
\usepackage{graphicx}
\usepackage{subfig}
\usepackage[top=1in, bottom=1in, left=1in, right=1in]{geometry}
\usepackage{enumitem}
\setlist[itemize]{leftmargin=*}
\usepackage{caption}
\allowdisplaybreaks

\title{CRKSPH-COMPATIBLE DISCRETIZATION OF \\
  THE SUPG AND SAAF TRANSPORT EQUATIONS}

\author{%
  \textbf{Brody R. Bassett$^1$ and J. Michael Owen$^1$} \\
  $^1$Lawrence Livermore National Laboratory  \\
  \ \ 7000 East Avenue, Livermore, CA, 94550 \\ 
\\
  \url{bassett4@llnl.gov}, \url{mikeowen@llnl.gov}
}
\newcommand{\authorHead}{Brody R. Bassett and J. Michael Owen}
\newcommand{\shortTitle}{CRKSPH-COMPATIBLE TRANSPORT}
\begin{document}
\maketitle
\justify 

\input{mc2021.tex}

\setlength{\baselineskip}{12pt}
\bibliographystyle{mc2021}
\bibliography{mc2021}

\end{document}

%% file: mc2021.tex
\section{INTRODUCTION}

Smoothed particle hydrodynamics (SPH) is a meshless approach to hydrodynamics
commonly used in astrophysics or for problems with unstable flows
\cite{monaghan2005smoothed}. There are several published methods
for doing radiation hydrodynamics using SPH with radiation diffusion
\cite{whitehouse2004smoothed,whitehouse2005faster,viau2006implicit,mayer2007fragmentation,petkova2009implementation},
including an implementation in the code used in this paper \cite{bassett2020efficient}.
The radiation transport equation has also been solved using collocation
methods \cite{sadat2006use,sadat2012meshless,kindelan2010application,liu2007least,zhao2013second,kashi2017mesh}
and variational forms \cite{liu2007meshless,bassett2019meshless}. 

Here we discretize the self-adjoint angular flux (SAAF) \cite{morel1999self}
and streamline-upwind Petrov-Galerkin (SUPG) \cite{pain2006streamline}
transport equations using reproducing kernels (RK) \cite{liu1995reproducing}
with the collocation method to produce a discretization that is compatible
with conservative reproducing kernel smoothed particle hydrodynamics
(CRKSPH) \cite{frontiere2017crksph} to permit future research into
meshless radiation hydrodynamics. The novelty of this approach is
a strong form discretization of the radiation transport equation that
does not involve explicit integration and includes kernels that can
reproduce functions up to an arbitrary polynomial order using reproducing
kernels. 

\section{MESHLESS KERNELS\label{sec:Meshless-kernels}}

\subsection{SPH Kernels}

A standard SPH kernel $W$ and its spatial derivatives can be written
in terms of a base kernel in reference space $W^{b}$ as
\begin{align}
W\left(x,H\right) & =W^{b}\left(\chi\left(\eta\left(x,H\right)\right)\right),\\
\partial_{x}^{\alpha}W & =\xi^{\alpha}\partial_{\chi}W^{b},\\
\partial_{x}^{\alpha\beta}W & =\frac{H^{\alpha\gamma}H^{\beta\gamma}-\xi^{\alpha}\xi^{\beta}}{\chi}\partial_{\chi}W+\xi^{\alpha}\xi^{\beta}\partial_{\chi,\chi}W,
\end{align}
with the transformed distance vector in ASPH space $\eta$, the scaled
distance $\chi$, and the normalized distance vector $\xi$,

\begin{gather}
\eta^{\alpha}=H^{\alpha\beta}x^{\beta},\\
\chi=\sqrt{\eta^{\alpha}\eta^{\alpha}},\\
\xi^{\alpha}=\frac{\eta^{\beta}}{\chi}H^{\alpha\beta}.
\end{gather}
The SPH kernel centered at the point $x_{j}$ and evaluated at the
point $x_{i}$ is written as
\begin{equation}
W_{ij}=W_{j}\left(x_{i}\right)=W\left(x_{i}-x_{j},H_{j}\right).
\end{equation}
SPH functions can be used to interpolate,
\begin{equation}
f_{\text{SPH}}\left(x\right)\approx\sum_{j}V_{j}f\left(x_{j}\right)W_{j}\left(x\right),
\end{equation}
where $V_{j}$ is the weight or volume of the kernel, but have a few
issues when used to discretize partial differential equations. The
interpolant cannot in general even reproduce a constant exactly, $\sum_{j}V_{j}W_{j}\left(x\right)\neq\text{const}$,
which reduces accuracy and leads to issues near boundaries. 

\subsection{RK Kernels}

To remedy these issues, the SPH functions can be augmented by RK,
which can interpolate up to the chosen order of polynomial exactly.
Defining $P\left(x\right)$ as the vector of polynomials with degree
less than or equal to the chosen order {[}or $P_{ij}=P\left(x_{i}-x_{j}\right)$
once evaluated{]} and $C_{i}$ as the corrections vector for the evaluation
point $i$, the RK functions and their first two derivatives are defined
as
\begin{align}
U_{ij} & =P_{ij}^{\top}C_{i}W_{ij},\\
\partial_{x_{i}}^{\alpha}U_{ij} & =\left(\partial_{x_{i}}^{\alpha}P_{ij}^{\top}C_{i}+P_{ij}\partial_{x_{i}}^{\alpha}C_{i}\right)W_{i}+P_{ij}^{\top}C\partial_{x_{i}}^{\alpha}W_{ij},\\
\partial_{x_{i}}^{\alpha\beta}U_{ij} & =\left(\partial_{x_{i}}^{\alpha\beta}P_{ij}^{\top}C_{i}+\partial_{x_{i}}^{\beta}P_{ij}^{\top}\partial_{x_{i}}^{\alpha}C_{i}+\partial_{x_{i}}^{\alpha}P_{ij}^{\top}\partial_{x_{i}}^{\beta}C_{i}+P_{ij}^{\top}\partial_{x_{i}}^{\alpha\beta}C_{i}\right)W_{ij}+P_{ij}^{\top}C_{i}\partial_{x_{i}}^{\alpha\beta}W_{ij}\nonumber \\
 & \qquad+\left(\partial_{x_{i}}^{\alpha}P_{ij}^{\top}C_{i}+P_{ij}^{\top}\partial_{x_{i}}^{\alpha}C_{i}\right)\partial_{x_{i}}^{\beta}W_{ij}+\left(\partial_{x_{i}}^{\beta}P_{ij}^{\top}C_{i}+P_{ij}^{\top}\partial_{x_{i}}^{\beta}C_{i}\right)\partial_{x_{i}}^{\alpha}W_{ij},
\end{align}
where $U_{ij}=U_{j}\left(x_{i}\right)$ is the kernel $U_{j}$ evaluated
at the point $x_{i}$. The corrections vectors are defined as
\begin{align}
C_{i} & =M_{i}^{-1}G,\\
\partial_{x_{i}}^{\alpha}C_{i} & =-M_{i}^{-1}\partial_{x_{i}}^{\alpha}M_{i}C_{i},\\
\partial_{x_{i}}^{\alpha\beta}C_{i} & =-M_{i}^{-1}\left(\partial_{x_{i}}^{\alpha\beta}M_{i}C_{i}+\partial_{x_{i}}^{\alpha}M_{i}\partial_{x_{i}}^{\beta}C_{i}+\partial_{x_{i}}^{\beta}M_{i}\partial_{x_{i}}^{\alpha}C_{i}\right),
\end{align}
where
\begin{align}
M_{i} & =\sum_{j}V_{j}P_{ij}P_{ij}^{\top}W_{ij},\\
\partial_{x_{i}}^{\alpha}M_{i} & =\sum_{j}V_{j}\left[\left(\partial_{x_{i}}^{\alpha}P_{ij}P_{ij}^{\top}+P_{ij}\partial_{x_{i}}^{\alpha}P_{ij}^{\top}\right)W_{ij}+P_{ij}P_{ij}^{\top}\partial_{x_{i}}^{\alpha}W_{ij}\right],\\
\partial_{x_{i}}^{\alpha\beta}M_{i} & =\sum_{j}V_{j}\left[\left(\partial_{x_{i}}^{\alpha\beta}P_{ij}P_{ij}^{\top}+\partial_{x_{i}}^{\alpha}P_{ij}\partial_{x_{i}}^{\beta}P_{ij}+\partial_{x_{i}}^{\beta}P_{ij}\partial_{x_{i}}^{\alpha}P_{ij}+P_{ij}\partial_{x_{i}}^{\alpha\beta}P_{ij}^{\top}\right)W_{ij}+P_{ij}P_{ij}^{\top}\partial_{x_{i}}^{\alpha\beta}W_{ij}\right.\nonumber \\
 & \qquad\left.+\left(\partial_{x_{i}}^{\alpha}P_{ij}P_{ij}^{\top}+P_{ij}\partial_{x_{i}}^{\alpha}P_{ij}\right)\partial_{x_{i}}^{\beta}W_{ij}+\left(\partial_{x_{i}}^{\beta}P_{ij}P_{ij}^{\top}+P_{ij}\partial_{x_{i}}^{\beta}P_{ij}\right)\partial_{x_{i}}^{\alpha}W_{ij}\right],
\end{align}
and $G=\left[1,0,0,\cdots\right]^{\top}$. For more information on
RK kernels, see Ref. \cite{liu1995reproducing}. The RK interpolant
is defined as
\begin{equation}
f_{\text{RK}}\left(x\right)=\sum_{j}V_{j}f\left(x_{j}\right)U_{j}\left(x\right).\label{eq:rk-interp}
\end{equation}
Letting $\mathbb{P}_{n}$ be the space of polynomials with degree
less than or equal to $n$, the RK kernels with correction order $n$
exactly interpolate any function in $\mathbb{P}_{n}$,
\[
f_{\text{RK}}\left(x\right)=f\left(x\right),\quad f\left(x\right)\in\mathbb{P}_{n}.
\]
The RK functions approximate delta functions ($U_{j}\left(x\right)\to\delta\left(x-x_{j}\right)$
as $H_{j}\to\infty$), so volume integrals of a smooth function and
the kernel can be approximated as 
\begin{equation}
\left\langle U_{i},f\right\rangle =\int_{V}U_{i}f\approx f_{i}.\label{eq:rk-eval}
\end{equation}
As only the values $f_{i}$ are known (and not their derivatives),
to calculate a derivative we first interpolate between the values
using Eq. (\ref{eq:rk-interp}) and then take derivatives of the interpolant,
\begin{equation}
\left\langle U_{i},\partial_{x}^{\alpha}f\right\rangle \approx\sum_{j}V_{j}f_{j}\left\langle U_{i},\partial_{x}^{\alpha}U_{j}\right\rangle \approx\sum_{j}V_{j}f_{j}\partial_{x_{i}}^{\alpha}U_{ji}.\label{eq:rk-deriv}
\end{equation}
Second derivatives are performed similarly,
\begin{equation}
\left\langle U_{i},\partial_{x}^{\alpha\beta}f\right\rangle \approx\sum_{j}V_{j}f_{j}\left\langle U_{i},\partial_{x}^{\alpha\beta}U_{j}\right\rangle \approx\sum_{j}V_{j}f_{j}\partial_{x_{i}}^{\alpha\beta}U_{ji}.\label{eq:rk-hess}
\end{equation}
For more complicated situations, as in Sec. \ref{subsec:SUPG-transport},
the the goal in deriving an RK derivative approximation is to isolate
$U_{i}$ so we can use Eq. (\ref{eq:rk-eval}) to evaluate the bilinear
integral of $U_{i}$ and whatever remains. 

We need one more derivative to do SAAF transport (Sec. \ref{subsec:SAAF-transport}),
which is the Hessian-like matrix $\partial^{\alpha}\left(g\partial^{\beta}f\right)$.
To derive this approximation, we want to avoid derivatives multiplied
by other derivatives, as then we would have to insert two interpolants
for $f$ and $g$, which would complicate the derivation and subsequent
implementation considerably. The approximation starts by writing two
different forms of the original derivative,
\begin{align}
\partial^{\alpha}\left(g\partial^{\beta}f\right) & =\partial^{\alpha\beta}\left(gf\right)-\partial^{\beta}g\partial^{\alpha}f-f\partial^{\alpha\beta}g,\\
\partial^{\alpha}\left(g\partial^{\beta}f\right) & =g\partial^{\alpha\beta}f+\partial^{\alpha}g\partial^{\beta}f.
\end{align}
Averaging these two approximations, we get 
\begin{equation}
\partial^{\alpha}\left(g\partial^{\beta}f\right)=\frac{1}{2}\left[\partial^{\alpha\beta}\left(gf\right)-f\partial^{\alpha\beta}g+g\partial^{\alpha\beta}f+\left(\partial^{\alpha}g\partial^{\beta}f-\partial^{\beta}g\partial^{\alpha}f\right)\right].
\end{equation}
If this identity is multiplied by the same vector twice, as it is
for SAAF, then this term can be symmetrized without affecting its
validity, since $k^{\alpha}k^{\beta}\partial^{\alpha}\left(g\partial^{\beta}f\right)=k^{\alpha}k^{\beta}\partial^{\beta}\left(g\partial^{\alpha}f\right)$.
Under symmetrization, the term involving derivatives on both $f$
and $g$ in parenthesis disappears,
\begin{equation}
k^{\alpha}k^{\beta}\partial^{\alpha}\left(g\partial^{\beta}f\right)=\frac{1}{2}k^{\alpha}k^{\beta}\left[\partial^{\alpha\beta}\left(gf\right)-f\partial^{\alpha\beta}g+g\partial^{\alpha\beta}f\right].
\end{equation}
The RK approximation of this derivative is 
\begin{align}
k^{\alpha}k^{\beta}\left\langle U_{i},\partial^{\alpha}\left(g\partial^{\beta}f\right)\right\rangle  & =\frac{1}{2}k^{\alpha}k^{\beta}\left[\left\langle U_{i},\partial^{\alpha\beta}\left(gf\right)\right\rangle -\left\langle U_{i},f\partial^{\alpha\beta}g\right\rangle +\left\langle U_{i},g\partial^{\alpha\beta}f\right\rangle \right]\nonumber \\
 & \approx\frac{1}{2}k^{\alpha}k^{\beta}\sum_{j}V_{j}\left(g_{j}f_{j}-g_{j}f_{i}+g_{i}f_{j}\right)\left\langle U_{i},\partial^{\alpha\beta}U_{j}\right\rangle \nonumber \\
 & \approx\frac{1}{2}k^{\alpha}k^{\beta}\sum_{j}V_{j}\left(g_{j}f_{j}-g_{j}f_{i}+g_{i}f_{j}\right)\partial_{x_{i}}^{\alpha\beta}U_{ji}\nonumber \\
 & =\frac{1}{2}k^{\alpha}k^{\beta}\sum_{j}V_{j}\left(g_{j}+g_{i}\right)\left(f_{j}-f_{i}\right)\partial_{x_{i}}^{\alpha\beta}U_{ji}.\label{eq:rk-saaf-deriv}
\end{align}
The extra $g_{i}f_{i}$ term that permits the factorization is equal
to zero since $\sum_{j}V_{j}\partial_{x_{i}}^{\alpha\beta}U_{ji}=0$.
As far as we are aware, this form of the RK derivative is novel. 

\section{DISCRETIZATION OF THE TRANSPORT EQUATION}

\subsection{SAAF Transport\label{subsec:SAAF-transport}}

The SAAF transport equation is
\begin{equation}
-\Omega^{\alpha}\Omega^{\beta}\partial_{x}^{\alpha}\left(\frac{1}{\sigma_{\tau}}\partial_{x}^{\beta}\psi\right)+\sigma_{\tau}\psi=s-\Omega^{\alpha}\partial_{x}^{\alpha}\left(\frac{s}{\sigma_{\tau}}\right),\label{eq:saaf-transport}
\end{equation}
with the source
\begin{equation}
s=\tau\psi^{n}+\frac{1}{4\pi}\sigma_{s}\phi+q,\label{eq:transport-source}
\end{equation}
where $\Omega$ is the streaming direction, $\sigma_{s}$ is the scattering
cross section, $\sigma_{a}$ is the absorption cross section, $\tau=1/\left(c\Delta t\right)$,
d $\sigma_{\tau}=\sigma_{s}+\sigma_{a}+\tau$, and $q$ is the nonhomogeneous
source. Equation \ref{eq:saaf-transport} is multiplied by $U_{i}$
and integrated,
\begin{equation}
\Omega^{\alpha}\Omega^{\beta}\left\langle U_{i},\partial_{x}^{\alpha}\left(\frac{1}{\sigma_{\tau}}\partial_{x}^{\beta}\psi\right)\right\rangle +\left\langle U_{i},\sigma_{\tau}\psi\right\rangle =\left\langle U_{i},s\right\rangle -\Omega^{\alpha}\left\langle U_{i},\partial_{x}^{\alpha}\left(\frac{s}{\sigma_{\tau}}\right)\right\rangle ,
\end{equation}
before the approximations in Eqs. (\ref{eq:rk-eval}), (\ref{eq:rk-deriv}),
and (\ref{eq:rk-saaf-deriv}) are inserted, 
\begin{equation}
-\Omega^{\alpha}\Omega^{\beta}\frac{1}{2}\sum_{j}V_{j}\left(\frac{1}{\sigma_{\tau,i}}+\frac{1}{\sigma_{\tau,j}}\right)\left(\psi_{j}-\psi_{i}\right)\partial^{\alpha\beta}U_{ji}+\sigma_{\tau,i}\psi_{i}=s_{i}-\Omega^{\alpha}\sum_{j}V_{j}\frac{s_{j}}{\sigma_{\tau,j}}\partial^{\alpha}U_{ji}.
\end{equation}

\subsection{SUPG Transport\label{subsec:SUPG-transport}}

The standard linear transport equation with backward Euler time differencing
can be written as 
\begin{equation}
\Omega^{\alpha}\partial_{x}^{\alpha}\psi_{m}+\sigma_{\tau}\psi=s,\label{eq:transport-equation}
\end{equation}
where $s$ is defined in Eq. (\ref{eq:transport-source}). To discretize
this equation with SUPG, Eq. (\ref{eq:transport-equation}) is multiplied
by $U_{i}+\kappa_{i}\Omega^{\alpha}\partial_{x}^{\alpha}U_{i}$ and
integrated, 
\begin{equation}
\Omega^{\alpha}\left\langle U_{i},\partial^{\alpha}\psi\right\rangle +\left\langle U_{i},\sigma_{\tau}\psi\right\rangle +\kappa_{i}\Omega^{\alpha}\Omega^{\beta}\left\langle \partial^{\beta}U_{i},\partial^{\alpha}\psi\right\rangle +\kappa_{i}\Omega^{\alpha}\left\langle \partial^{\alpha}U_{i},\sigma_{\tau}\psi\right\rangle =\left\langle U_{i},s\right\rangle +\kappa_{i}\Omega^{\alpha}\left\langle \partial^{\alpha}U_{i},s\right\rangle .
\end{equation}
The derivatives are moved away from the $U_{i}$ terms through integration
by parts and the surface integrals are discarded, 
\begin{equation}
\Omega^{\alpha}\left\langle U_{i},\partial^{\alpha}\psi\right\rangle +\left\langle U_{i},\sigma_{\tau}\psi\right\rangle -\kappa_{i}\Omega^{\alpha}\Omega^{\beta}\left\langle U_{i},\partial^{\alpha\beta}\psi\right\rangle -\kappa_{i}\Omega^{\alpha}\left\langle U_{i},\partial^{\alpha}\left(\sigma_{\tau}\psi\right)\right\rangle =\left\langle U_{i},s\right\rangle -\kappa_{i}\Omega^{\alpha}\left\langle U_{i},\partial^{\alpha}s\right\rangle .
\end{equation}
Finally, the integrals are performed as in Eqs. (\ref{eq:rk-eval}),
(\ref{eq:rk-deriv}), and (\ref{eq:rk-hess}) to get the discretized
equation,
\begin{equation}
\sum_{j}V_{j}\left[\left(1-\kappa_{i}\sigma_{\tau,j}\right)\Omega^{\alpha}\partial^{\alpha}U_{ji}-\kappa_{i}\Omega^{\alpha}\Omega^{\beta}\partial^{\alpha\beta}U_{ji}\right]\psi_{j}+\sigma_{\tau,i}\psi_{i}=s_{i}-\kappa_{i}\Omega^{\alpha}\sum_{j}V_{j}s_{j}\partial^{\alpha}U_{ji}.
\end{equation}
The stabilization coefficient $\kappa$ is set to be the average distance
between neighboring points, which for a function like the Gaussian
means that the relative magnitudes of $U_{i}$ and $\kappa_{i}\partial_{x}^{\alpha}U_{i}$
do not depend on the smoothing length $H$. 

It is of note that for constant cross sections and $\kappa=1/\sigma_{\tau}$
(where the effective mean free path is equal to the point spacing),
the discretized SUPG and SAAF equations are exactly equal. If $\kappa$
were taken to be spatially-dependent in the SUPG test function, the
resulting equation is
\begin{multline}
\sum_{j}V_{j}\left[\left(1-\kappa_{j}\sigma_{\tau,j}\right)\Omega^{\alpha}\partial^{\alpha}U_{ji}\psi_{j}-\frac{1}{2}\Omega^{\alpha}\Omega^{\beta}\left(\kappa_{j}+\kappa_{i}\right)\left(\psi_{j}-\psi_{i}\right)\partial^{\alpha\beta}U_{ji}\right]+\sigma_{\tau,i}\psi_{i}\\
=s_{i}-\Omega^{\alpha}\sum_{j}V_{j}\kappa_{j}s_{j}\partial^{\alpha}U_{ji},
\end{multline}
which is exactly equal to the SAAF equation even with spatially-dependent
cross sections if $\kappa=\sigma_{\tau}$. In practice, this equation
does not perform as well as a constant $\kappa_{i}$. For a constant
$\kappa_{i}$ and a Gaussian-like kernel, the stabilization can be
interpreted as upwinding the kernel. If $\kappa$ is dependent on
$\sigma_{\tau}$ or is spatially-dependent within a test function,
then the shape of the test function will change based on the extent
of other kernels or refinement in space or time. 

\section{RESULTS}

We consider three test cases, a purely-absorbing slab in 1D with an
analytic solution and steady-state and time-dependent manufactured
solutions in 1D and 2D. Second-order RK corrections (e.g. in 2D a
polynomial vector of $P^{\top}=[1,x,y,x^{2},xy,y^{2}${]}), a kernel
support of six times the point spacing (see Ref. \cite{owen2010asph}
for details), and Wendland 33 kernels \cite{wendland1995piecewise}
are used for all cases. 

\begin{figure}
\begin{centering}
\includegraphics[width=0.47\textwidth]{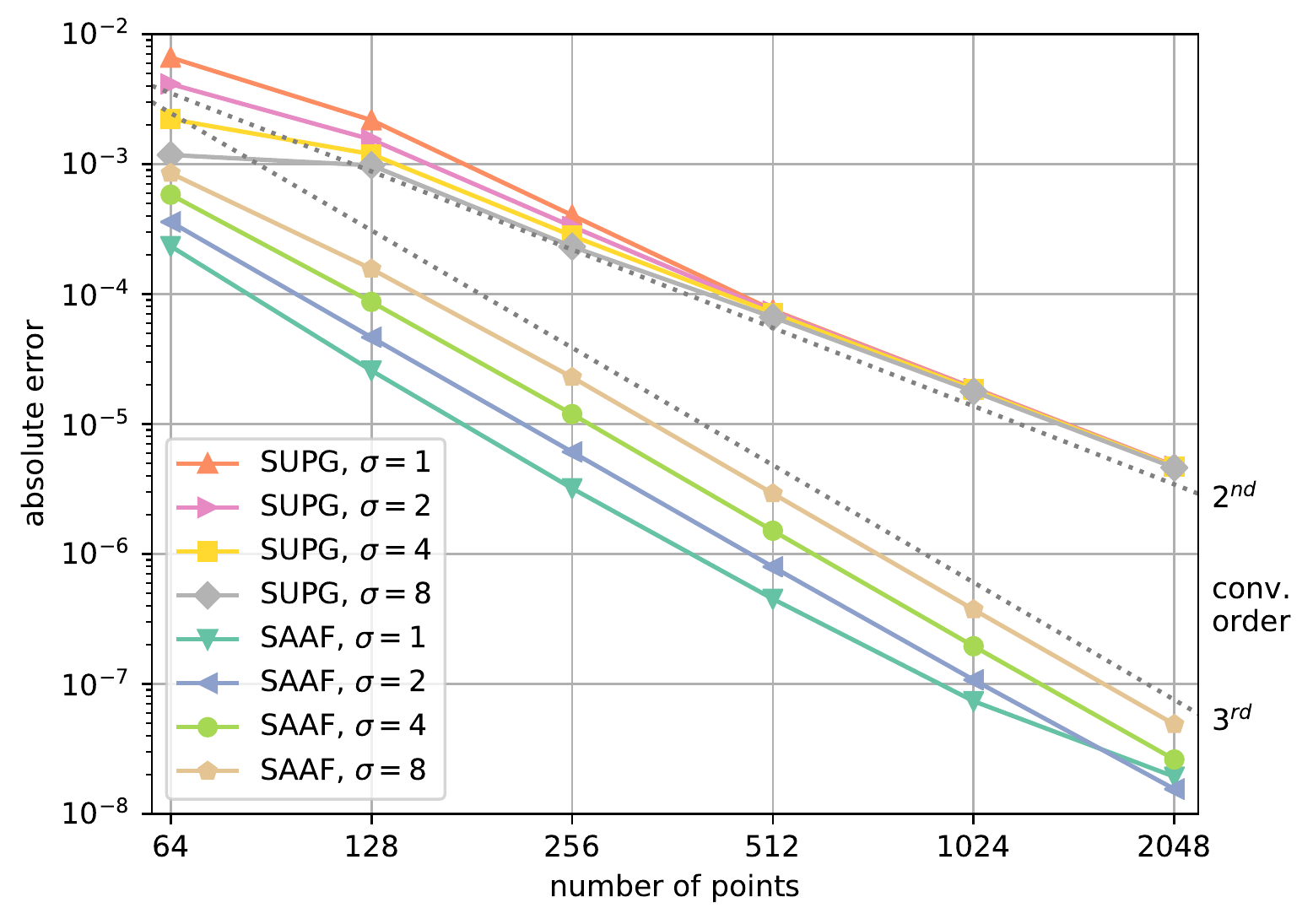}
\par\end{centering}
\caption{Absolute error for the purely absorbing problem for SAAF and SUPG
for four values of the absorption cross section.}
\label{fig:purely-absorbing-error}
\end{figure}

\subsection{Purely Absorbing}

The steady-state purely absorbing problem involves setting a boundary
value for the angular flux, $\psi_{0}$, incident on a purely-absorbing
slab with an absorption cross section of $\sigma_{a}$. Given the
distance from the boundary $x$ and the direction cosine $\mu=\Omega_{x}$,
the analytic solution to this problem is 
\[
\psi=\psi_{0}\exp\left(-\frac{\sigma_{a}x}{\mu}\right).
\]
We consider each of SAAF and SUPG for four cross section values, $\sigma_{a}=1,2,4,8$,
and calculate the absolute error of the numeric solution compared
to the analytic, 
\[
\epsilon_{L_{1}}=\int_{V}\left|\psi_{\text{num}}-\psi_{\text{ana}}\right|\approx\sum_{i}V_{i}\left|\psi_{\text{num},i}-\psi_{\text{ana},i}\right|.
\]
Figure \ref{fig:purely-absorbing-error} shows the error for all eight
cases. As expected, the cases with higher cross sections have larger
error as the density of points needed to resolve the steep solution
is larger. The SUPG and SAAF results are consistent with second and
third-order converge, respectively. In the next problem, the convergence
orders swap, but it is unclear from the results that we have why that
is. Possible reasons could be the difference in boundary and internal
source treatments between SAAF and SUPG or a preference for SAAF toward
problems with large gradients. 

\subsection{Steady-State Manufactured}

This problem is a steady-state manufactured solution,
\begin{equation}
\psi_{\text{man}}=1.2+\prod_{\alpha=1}^{\text{dimension}}\cos\left(2\pi x^{\alpha}\right),
\end{equation}
for each of SAAF and SUPG in 1D and 2D. This solution is inserted
into the steady-state first-order transport equation {[}Eq. (\ref{eq:transport-equation})
with $\tau\to0${]} and solved for $q$, after which this $q$ is
used in a transport calculation to calculate a numeric flux, $\psi_{\text{numeric}}$.
We then calculate the volume-integrated relative $L_{2}$ error between
the manufactured and numeric solutions,
\[
\epsilon_{L_{2},\text{rel}}=\frac{\int_{V}\left|\psi_{\text{num}}-\psi_{\text{man}}\right|}{\int_{V}\psi_{\text{man}}}\approx\frac{\sum_{i}V_{i}\left|\psi_{\text{num},i}-\psi_{\text{man},i}\right|}{\sum_{i}V_{i}\psi_{\text{man},i}}.
\]
The points are placed uniformly in one case and are randomly perturbed
by up to 0.2 times the point spacing in the other. 

\begin{figure}
\begin{centering}
\subfloat[Uniform point distribution.]{\includegraphics[width=0.47\textwidth]{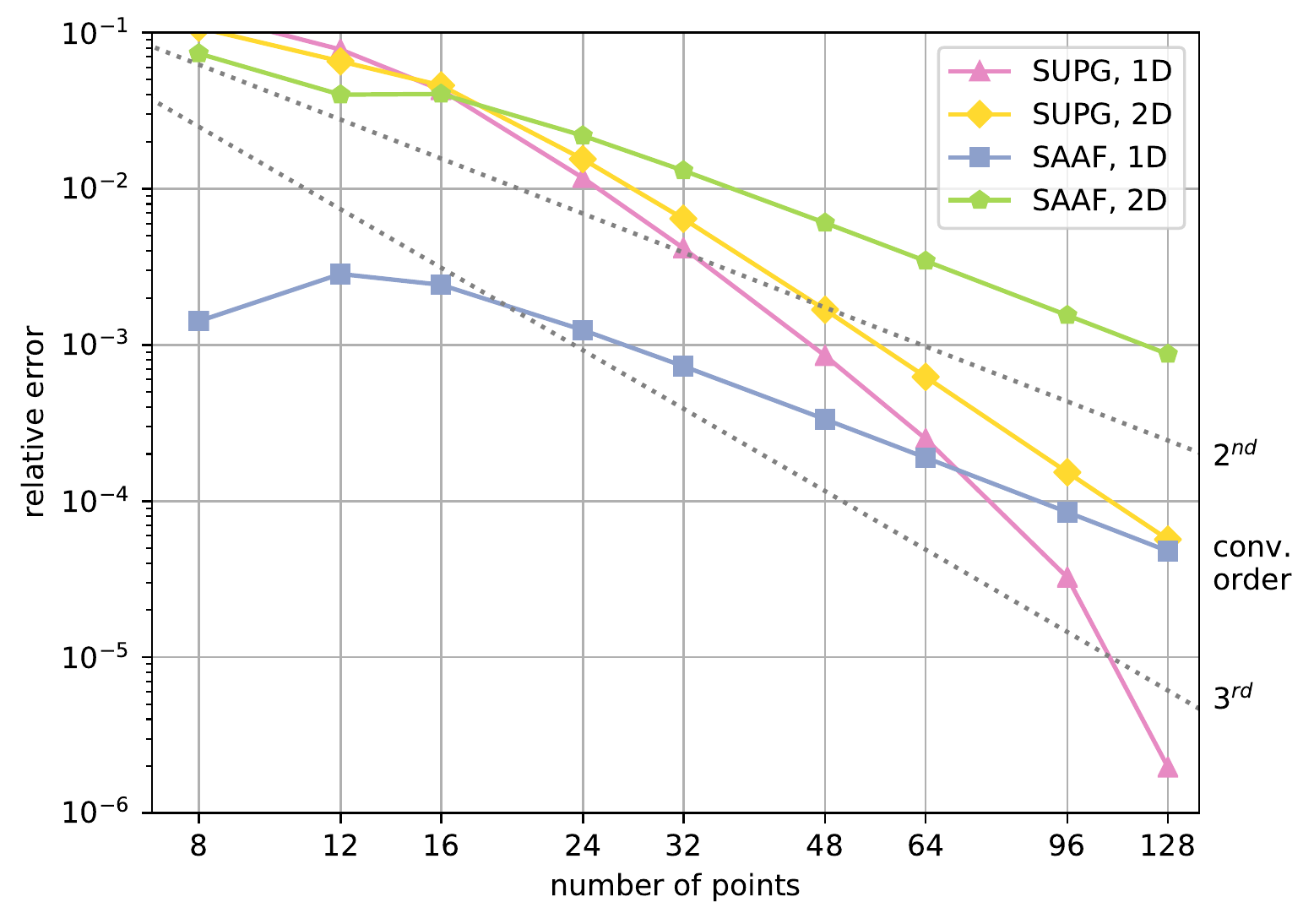}

}\subfloat[Non-uniform point distribution.]{\begin{raggedleft}
\includegraphics[width=0.47\textwidth]{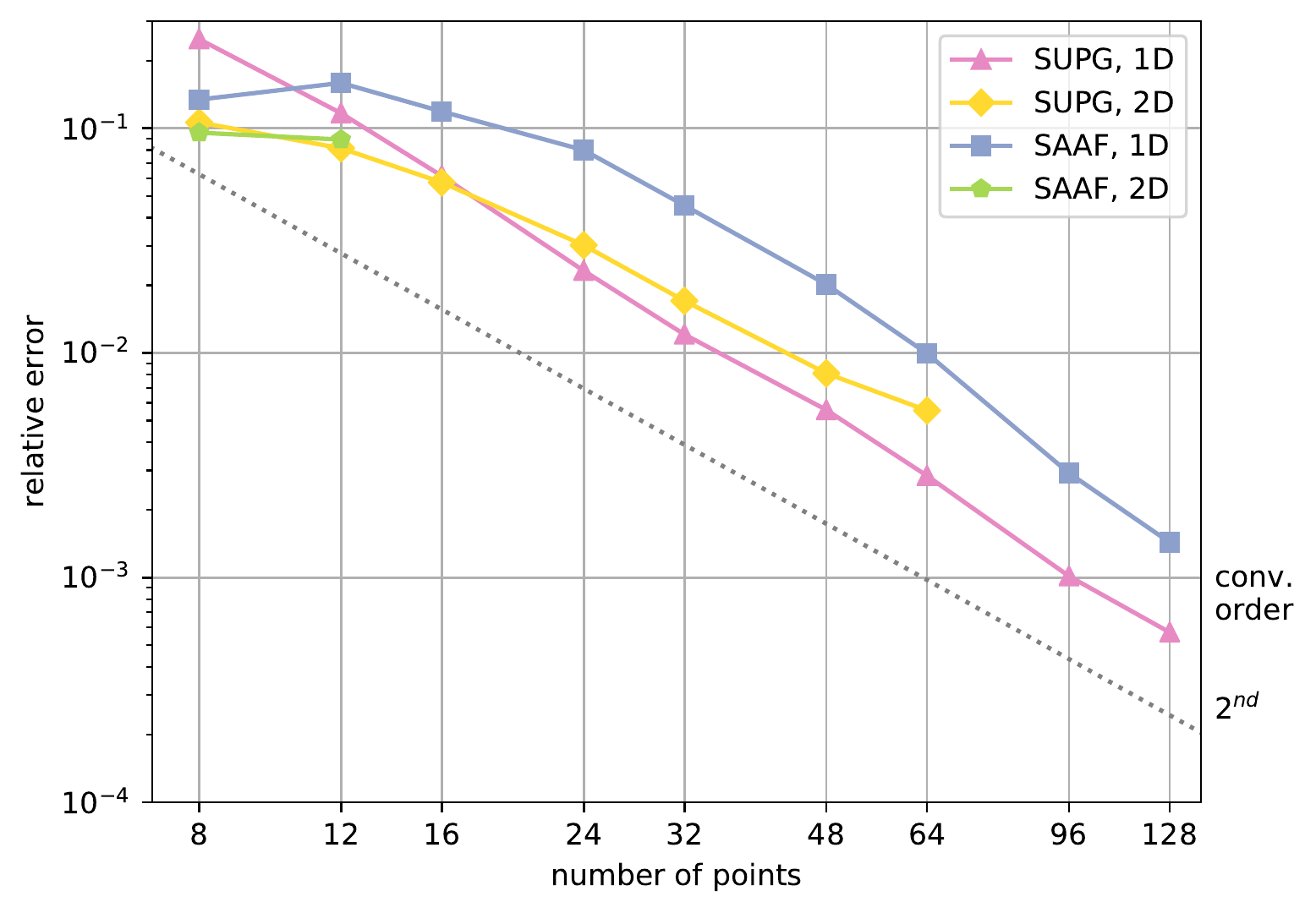}
\par\end{raggedleft}
}
\par\end{centering}
\caption{Relative error for the steady-state sinusoidal manufactured solution
for SAAF and SUPG in 1D and 2D. The points that are missing in the
non-uniform case represent simulations that did not converge. }
\label{fig:sinusoidal-error}
\end{figure}

Figure \ref{fig:sinusoidal-error} shows the error for the uniform
and non-uniform cases. For uniform points, the SAAF transport converges
with second-order accuracy. The SUPG transport convergence with third-order
accuracy in 2D and spectral accuracy in 1D. For non-uniform points,
the GMRES transport solver with ILUT preconditioning in the more refined
SUPG cases and most of the SAAF cases did not converge to the specified
tolerance of $10^{-14}$ within 1,000 iterations, so these points
are not included on the plot. The SUPG results in 1D and 2D as well
as the SAAF results in 1D imply second-order convergence. 

The SAAF results in 2D are particularly sensitive to random pertubations
in point locations and do not converge for a majority of the cases
with randomized point locations. The SUPG results in 2D perform better
but still struggle to converge using Trilinos GMRES solver with ILUT
preconditioning \cite{heroux2005overview}. For problems in which
the point locations are not chosen (e.g. in a simulation coupled with
hydrodynamics), this could present issues. When using a direct solver,
the results continue to converge, but this is impractical to do for
realistic problems. General-purpose AMG preconditioners, such as BoomerAMG
in Hypre \cite{falgout2002hypre}, fare worse than the ILUT preconditioners.
Because of this, additional work needs to be done on selecting proper
solvers for the equations; special-purpose preconditioners such as
the transport solver pAIR described in Ref. \cite{hanophy2020parallel}
may help. 

\begin{figure}
\begin{centering}
\subfloat[Uniform point distribution.]{\includegraphics[width=0.47\textwidth]{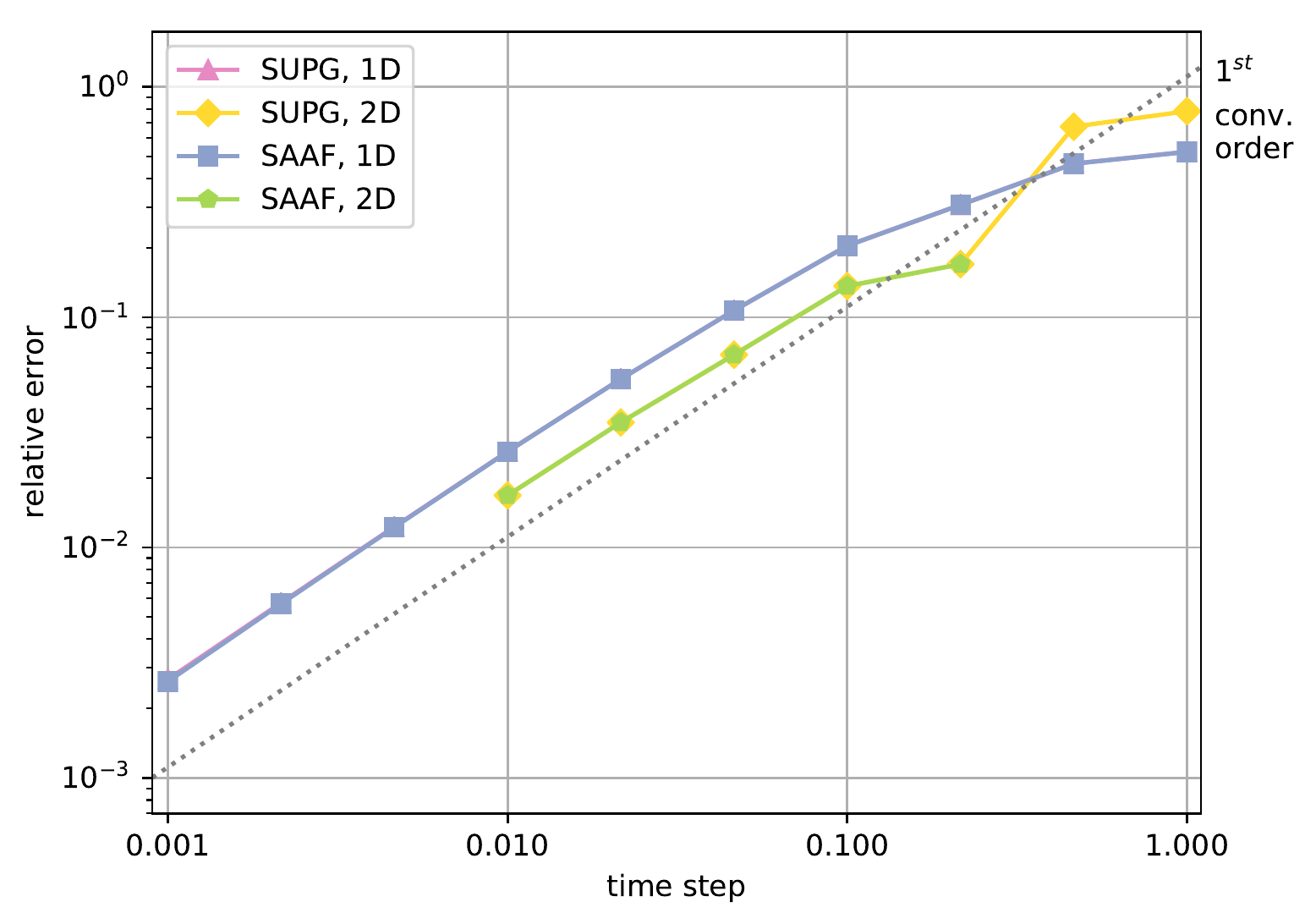}

}\subfloat[Non-uniform point distribution.]{\begin{raggedleft}
\includegraphics[width=0.47\textwidth]{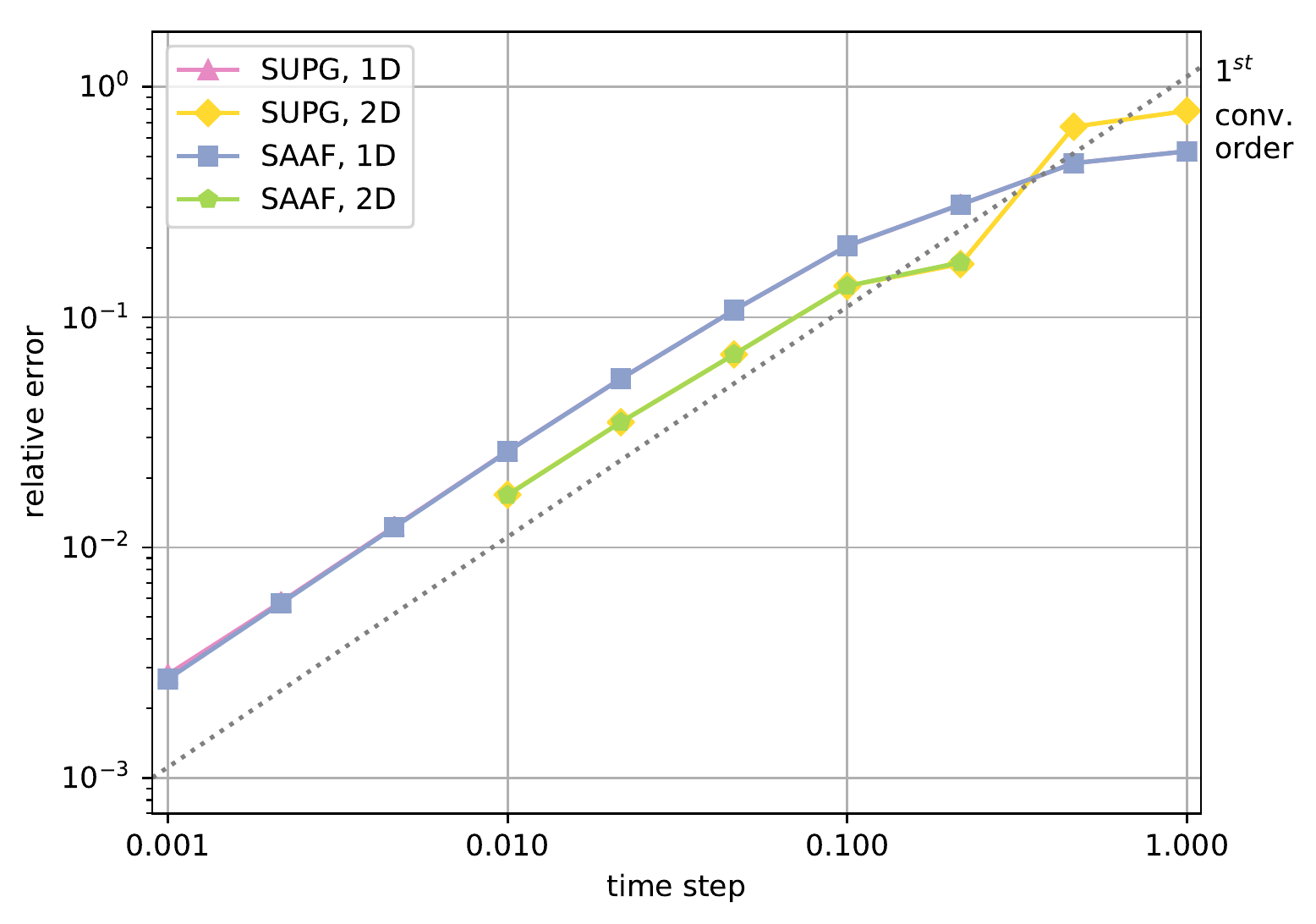}
\par\end{raggedleft}
}
\par\end{centering}
\caption{Relative error for the time-dependent sinusoidal manufactured solution
for SAAF and SUPG in 1D and 2D. The SAAF results for the two largest
time steps in 2D did not converge and are excluded.}
\label{fig:sinusoidal-time-error}
\end{figure}

\subsection{Time-Dependent Manufactured}

This problem is similar to the steady-state version except with a
time-dependent term added into the manufactured solution,
\[
\psi_{\text{man}}=1.2+\prod_{\alpha=1}^{\text{dimension}}\cos\left(2\pi\left(x^{\alpha}+t\right)\right),
\]
that also results in a time-dependent source. The number of points
is held constant at 64 along each dimension. The time step is chosen
and held constant throughout the simulation. As before, the points
are held constant for one set of tests and randomly perturbed by up
to 0.2 times the point spacing in the other set. The error is calculated
as in the previous test. 

Figure \ref{fig:sinusoidal-time-error} shows the convergence results.
There is less than one percent difference between the solutions with
randomized and non-randomized points, which indicates that the error
in the time-dependent case is dominated by the time step. Both the
uniform and randomized sets of tests are consistent with first-order
convergence in time, as expected with our fully-implicit time discretization.
The solver struggles to converge for SAAF when the time step is large.
See the steady-state section for a discussion on preconditioners and
convergence. 

\section{CONCLUSIONS}

The SAAF and SUPG transport equations are discretized using RK with
a combination of familiar derivatives and a novel second derivative.
The resulting equations involve only evaluations of kernels and physical
data at the nodal centers. With second-order RK kernels, the uniform
and non-uniform results are consistent with at least second-order
convergence. The SAAF results for non-uniform points struggle to converge
in 2D. While some of these problems are allayed by the effective absorption
in the time-dependent problem, the SAAF and SUPG solvers may benefit
from specialized preconditioners. In order to extend the results to
higher order, a larger kernel support would be needed that would exacerbate
these issues. Once the solvers are appropriately preconditioned, this
method would be attractive for coupling to an SPH or CRKSPH simulation.
The discretization is similar and the evaluations require no more
data than the original SPH calculation except the cross sections at
the evaluation nodes. 

\section*{ACKNOWLEDGEMENTS}

This work was performed under the auspices of the U.S. Department
of Energy by Lawrence Livermore National Laboratory under Contract
DE-AC52-07NA27344. LLNL-CONF-819823.